\pdfoutput=1

\documentclass[11pt]{article}

\usepackage[final]{acl}

\usepackage{times}
\usepackage{latexsym}

\usepackage[T1]{fontenc}

\usepackage[utf8]{inputenc}

\usepackage{microtype}

\usepackage{inconsolata}

\usepackage{graphicx}

\usepackage{amsmath}
\usepackage{amsfonts}

\usepackage[ruled,vlined,linesnumbered]{algorithm2e}
\usepackage{booktabs}
\usepackage{multicol}
\usepackage{multirow}
\usepackage{url}

\title{Exploiting Prefix-Tree in Structured Output Interfaces for Enhancing Jailbreak Attacking}

\author{
    Yanzeng Li\textsuperscript{1,2}\thanks{\hspace{0.15cm}This work was done during an internship at Tencent.}~,
    Yunfan Xiong\textsuperscript{1},
    Jialun Zhong\textsuperscript{1},
    Jinchao Zhang\textsuperscript{2},
    Jie Zhou\textsuperscript{2},
    Lei Zou\textsuperscript{1}\thanks{~~Corresponding author} \\
    $^{1}$Wangxuan Institute of Computer Technology, Peking University \\
    $^{2}$Pattern Recognition Center, WeChat AI, Tencent Inc. 
}

\begin{document}
\maketitle

\begin{abstract}
\textcolor{red}{Content warning: This paper contains content generated by LLMs that may be offensive or harmful, discretion is recommended.}

The rise of Large Language Models (LLMs) has led to significant applications but also introduced serious security threats, particularly from jailbreak attacks that manipulate output generation. These attacks utilize prompt engineering and logit manipulation to steer models toward harmful content, prompting LLM providers to implement filtering and safety alignment strategies. 
We investigate LLMs’ safety mechanisms and their recent applications, revealing a new threat model targeting structured output interfaces, which enable attackers to manipulate the inner logit during LLM generation, requiring only API access permissions.
To demonstrate this threat model, we introduce a black-box attack framework called AttackPrefixTree (APT). APT exploits structured output interfaces to dynamically construct attack patterns. By leveraging prefixes of models' safety refusal response and latent harmful outputs, APT effectively bypasses safety measures. 
Experiments on benchmark datasets indicate that this approach achieves higher attack success rate than existing methods. 
This work highlights the urgent need for LLM providers to enhance security protocols to address vulnerabilities arising from the interaction between safety patterns and structured outputs.~\footnote{The code is available at \url{https://github.com/lsvih/attackPrefixTree}}
\end{abstract}

\section{Introduction}
\label{sec:intro}

With the rapid development of Large Language Models~(LLMs) in recent years and their widespread application in various domains~\cite{minaee2024large,zhao2023survey,hadi2023survey}, the emergence of attacks targeting these models, particularly \textbf{jailbreak} attacks aimed at manipulating output generation, poses significant security threats~\cite{shen2024anything, yu2023gptfuzzer, yi2024jailbreak}. 
Such attacks primarily leverage techniques like prompt engineering to inject harmful inputs prior to generation, guiding the model toward producing harmful, false, or unethical outputs. 
In response, LLM researchers and vendors employ various safety mechanisms to mitigate the risk of generating harmful content, including output filtering, and constrained decoding~\cite{shayegani2023survey, chua2024ai}. 
Specifically, post-training methods, such as Safety Alignment~\cite{ji2024pku, ji2024beavertails}, involve constructing safety-aligned datasets to guide the model in avoiding potential harmful patterns, prioritizing safe and benign responses over harmful ones in the output rankings.

As LLMs continue to evolve in domain-specific applications, structured output has emerged as a critical feature~\cite{liu2024we,beurer2024guiding}. By fine-tuning the capacity for structured output, developing structured generation templates, and controlling model logit masks during generation through automated mechanisms, LLMs can produce outputs that conform to various structured formats such as JSON and YAML~
\cite{zheng2024sglang}. 
However, these capabilities also introduce new potential security vulnerabilities, as attackers may gain direct or indirect control over the decoding process, thereby influencing model outputs.

In this paper, we conduct a comprehensive analysis of the current state of research on safety fine-tuning and structured output. We find that while existing protective measures can resist straightforward attacks to some extent, the continual evolution of attack techniques poses a growing challenge to their effectiveness. Our motivations can be summarized as following observations:

    \noindent\textit{\textbf{Observation 1}}: During safety alignment, LLMs significantly reduce the probability of predicting and sampling harmful content through post-training; however, this does not lower the probability of such harmful outputs below the average token prediction probability. This implies that by expanding the sampling and output range, these harmful outputs are still obtainable~\cite{yuan2024refuse, xie2024sorry}.
    
    \noindent\textit{\textbf{Observation 2}}: During post-training, the safe responses of the model, such as generating phrases such as ``I am sorry, I cannot produce unethical content'', often depend on the safety alignment dataset used. To ensure the stability of the model's training and user experience, these safety patterns are typically at sentence level and are drawn from a limited set~\cite{li2024lockpicking}.
    
    \noindent\textit{\textbf{Observation 3}}: In model inference, the token-by-token prediction process of LLMs strives to maintain the coherence and completeness of sentences. This means that once the prefix tokens of a generated sentence are determined, it is challenging for the model to alter the overall semantics of the sentence through safety measures~\cite{zhang2024jailbreak}.

Thus, the token-by-token inference mode of LLMs conflicts with the sentence-level safety patterns learned during post-training, creating opportunities for attackers to manipulate subsequent output by intervening in the prefix of the model's output. Moreover, structured output provides attackers with mechanisms to intervene in the model's logit output, allowing for potential control over the prefix sampling results to generate harmful and unethical responses.
This attack paradigm inspires the proposal of a novel framework, named AttackPrefixTree (APT). By utilizing service interfaces that provide structured outputs, we online construct a tree structure that includes model safety pattern prefixes and harmful outputs. Through iterative calling model and sampling, the APT continually expands nodes representing safe prefixes and harmful outputs, ultimately guiding the formation of harmful outputs and the potential security measure prefix tree of the model. 
Our investigation reveals that existing models struggle to address this dynamic form of attack under the proposed framework, indicating that model service providers must carefully reconsider and enhance security measures when offering structured output interfaces. The contributions of our proposed method can be summarized as follows:
\begin{enumerate}
    \item We propose a new threat model for jailbreak attacks that exploits structured output interfaces, demonstrating how adversaries can bypass sentence-level refusal patterns through token-level manipulation of safety prefixes. Our analysis reveals fundamental vulnerabilities in current safety alignment approaches when confronted with structured output constraints.
    \item We implement the proposed threat model by designing AttackPrefixTree (APT), a black-box attack framework that systematically combines online tree-based exploration of safety prefixes with constrained decoding to dynamically suppress refusal patterns. By leveraging structured output interfaces, our approach enables token-level sampling manipulation without requiring model weight access while automatically retrieving refusal patterns.
    \item Extensive evaluations across JailBreakBench, AdvBench, and HarmBench demonstrate state-of-the-art attack success rates comparing with existing methods, suggesting that model providers reconsider the security threats associated with open access to direct logit manipulation interfaces. We further discover the implications of this threat model on the reasoning models, revealing new potential risks in multi-stage reasoning model generation processing.
\end{enumerate}

\section{Preliminary \& Related Works}

\subsection{Structured Output}\label{sec:output}


\citet{openai_structured_outputs} introduced structured output in the API model, which made the outputs adhere reliably to the developer-supplied JSON schemas. A widely used method is constrained decoding. Outlines \citep{willard2023efficient} and LM Format Enforcer\footnote{\url{https://github.com/noamgat/lm-format-enforcer}} are famous libraries that enforce the output format (JSON Schema, Regex, etc.) of a language model, used by vLLM \citep{kwon2023efficient} as the structure output backend. 

\begin{figure}[ht]
    \centering
    \includegraphics[width=\linewidth]{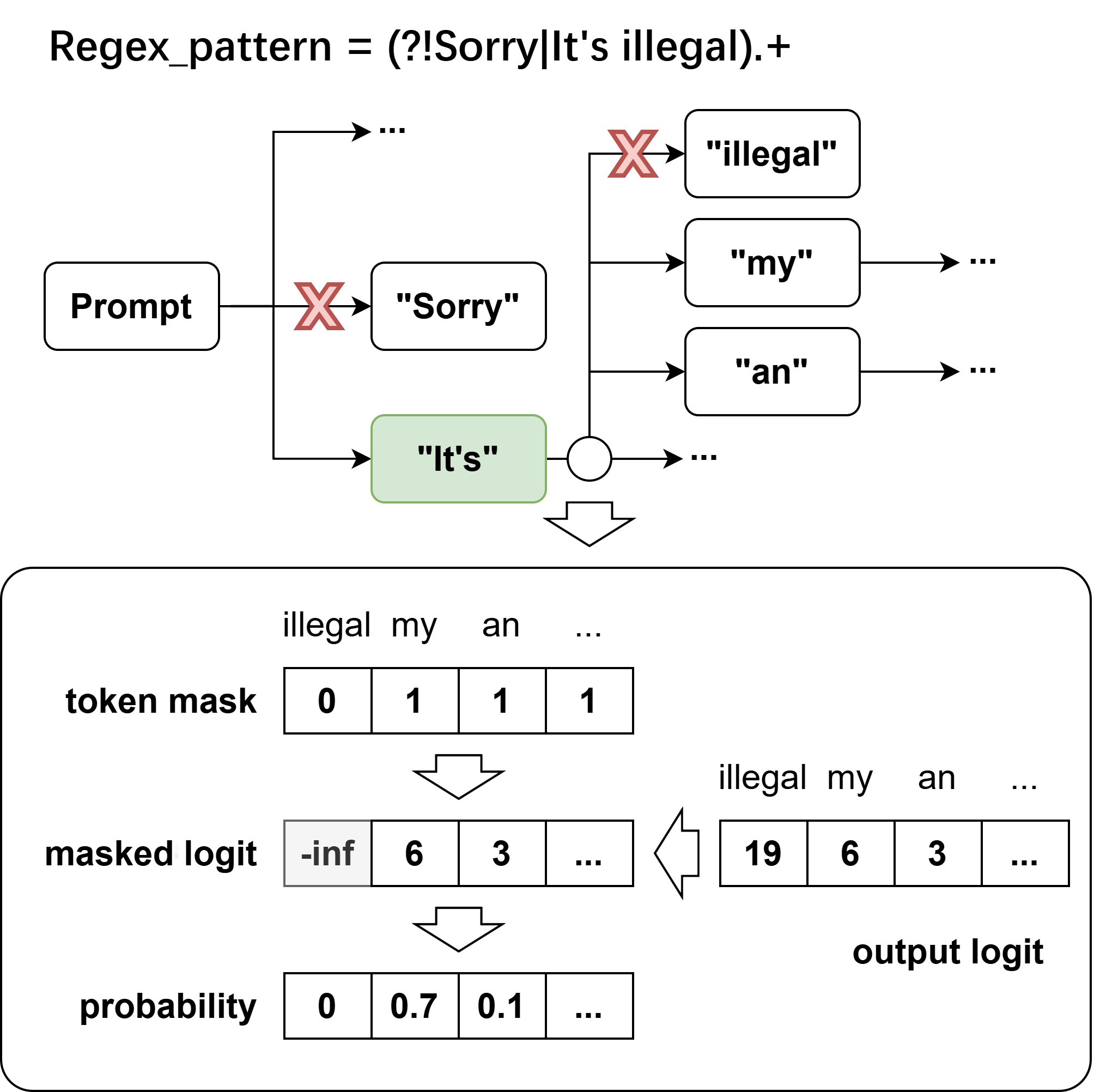}
    \caption{An example of constrained decoding. The constrained pattern in Regex is \texttt{(?!Sorry|It's illegal).*}. }
    \label{fig:constrained_output}
\end{figure}

During the LLM generation phase, constrained decoding dynamically maintains a valid token mask corresponding to predefined restrictions. This mask is applied to modify the output logits of the LLM, thereby generating a constrained probability distribution from which the next token is sampled, which ensures that each generated token will obey the given constraint. As illustrated in Figure~\ref{fig:constrained_output}, this approach enforces a regex-based constraint, \texttt{(?!Sorry|It's illegal).*}, which prohibits the generated text from beginning with phrases such as ``Sorry'' or ``It's illegal''.

Outlines, LM Format Enforcer, and instructor\footnote{\url{https://github.com/instructor-ai/instructor}} are famous structured output repositories for large language models. Most of them build a finite-state auto-machine to maintain the valid tokens of current state determined by previous tokens. The examples in Outlines used context-free grammar to support more features of the JSON schema based on Lark\footnote{\url{https://github.com/lark-parser/lark}}. 


\subsection{LLM Jailbreak}

Jailbreak attacks aim to surpass LLMs' safety mechanisms, coercing models to generate harmful content like ``How to damage a traffic light''. \citet{DBLP:conf/nips/0001HS23} introduces the concept of \textit{jailbreak attack} that is an attempt to elicit a direct response to a harmful prompt for by prompt modification, a problem which has received widespread attention. The possibility of jailbreak comes from models' \textit{Competing Objectives}, i.e., inconsistencies objective of helpfulness and harmlessness during alignment and mismached objective of pre-training and safe-alignment training. The success of a jailbreak attack often hinges on initial response tokens, i.e., opening phases like ``Sorry'' or ``It's illegal'' would trigger LLMs' refusal to provide answer, while prefixes such as ``Sure'' and ``Here is'' usually signify a compliance response, which could come a successful attack. The jailbreak approaches can be categorized into white- and black-box methods based on whether the attacker necessitates access to the models' weights.



\textbf{White-box} attacks leverage model internals. Prompt injection techniques \citep{DBLP:journals/corr/abs-2307-15043, DBLP:journals/corr/abs-2404-07921} optimize adversarial suffixes via gradient-based methods, while \citet{zhu2024advprefixobjectivenuancedllm} expands this to multi-prefix optimization. Other approaches directly manipulate logits: \citet{DBLP:journals/corr/abs-2405-13068} masks refusal-related tokens, and \citet{DBLP:conf/emnlp/LiuZHLZS24} refines probability distributions during decoding to suppress safety disclaimers.




\textbf{Black-box.} In scenarios where access to the model weights is unavailable, we are limited to employing prompt design strategies to attempt jailbreak attacks. 
The black-box attacks usually rely on iterative prompt engineering. Recently studies~\citep{anil2024manyshot,DBLP:journals/corr/abs-2404-01833,DBLP:journals/corr/abs-2408-11313,DBLP:journals/corr/abs-2408-04686,DBLP:journals/corr/abs-2405-05610} typically use multiple iterations to analyze and refine the prompt automatically, tile the attack successes. 


\subsubsection{Auto-method}

The white-box methods need the access permission to model's weight and usually need lots of computation and times. The black-box are ad-hoc methods, depend on human's carefully design and many tires to find a new pattern. In many situations, we do not have access to the model's weight, and spending time designing a jailbreak attack prompt is unaffordable. Methods for automatically generating jailbreaks are proposed (e.g. \citep{DBLP:journals/corr/abs-2310-08419,DBLP:journals/corr/abs-2311-03348,DBLP:journals/corr/abs-2309-10253}). Most methodologies employ various strategies to generate a substantial pool of candidate prompts (predefined categories, via iterative model self-improvement, etc.). These approaches are complemented by a discriminator that assesses the effectiveness of each candidate prompt in successfully executing an attack. By navigating through a defined search space and employing a search-and-validate mechanism, these methods autonomously identify prompts capable of achieving successful attacks.

\citet{DBLP:conf/iclr/HuangGXL024} found that open source safety-aligned LLMs are still vulnerable to jailbreak attack. The alignment procedures are based on a subset of decoding settings while remaining vulnerable to attacks under other decoding settings, such as different temperature, decoding penalty, and constraint decoding.  

\citet{DBLP:conf/acl/ZhangGZC00CW24} investigates the impact of constrained decoding on jailbreak attacks and proposes a novel attack method termed EnDec. EnDec involves training an auxiliary model designed to detect whether the current output contains negative phrases such as "Sorry" or "illegal". Upon detection of such terms, the method forcibly replaces the token with its antonym, thereby circumventing the constraints imposed by the decoding process.


\subsubsection{Anti-Jailbreak Techniques}

Significant research efforts have been dedicated to developing anti-jailbreak strategies to counteract malicious attacks on LLMs. These methods typically target specific jailbreak vulnerabilities and can be categorized into four primary approaches:

\begin{enumerate}
    \item \textbf{Safety Alignment}: These techniques focus on identifying harmful examples in training data or model outputs and refining the model's safety through alignment processes. Representative work includes post-training alignment optimizations and safety-enhanced fine-tuning \cite{haider2024phi3safetyposttrainingaligning, DBLP:journals/corr/abs-2406-01288}.
    \item \textbf{Prompt Perturbation}: This strategy involves systematically modifying user prompts to neutralize potential risks, such as adding semantic constraints or rephrasing queries to prevent unsafe responses \citep{DBLP:journals/corr/abs-2408-08924, DBLP:journals/corr/abs-2402-16192}.
    \item  \textbf{Input/Output Detection}: Researchers have developed auxiliary detection models that screen both user inputs and LLM outputs for harmful content, enabling real-time blocking of unsafe interactions \citep{wang-etal-2024-self, DBLP:conf/emnlp/LiuXWS24}.
    \item \textbf{Parameter Intervention}: Advanced methods directly analyze neural weights to identify safety-critical parameters, then enhance robustness through selective parameter modification or activation steering \citep{DBLP:journals/corr/abs-2406-11717, DBLP:conf/icml/WeiHHXQXMW024, DBLP:journals/corr/abs-2410-10343}.
\end{enumerate}





\section{Methodology}

\subsection{Threat Model}

We consider a black-box attack scenario where an attacker exploits the LLM’s \textit{structured output} functionality (e.g., regex constraints, JSON/XML formatting) to bypass safety guardrails. The attacker interacts solely via the LLM’s public API and has no access to model parameters, gradients, or internal safety mechanisms. The attack framework is defined as follows:

\textbf{Attack Strategy}: The attacker crafts (1) a malicious query $ P_q $ (e.g., ``How to damage traffic light''), and (2) a \textit{structured output pattern} $ P_p $ (e.g., regex ``(?!Sorry|I cannot assist)''). These components are combined into a jailbreak template $ P_J $:
\begin{equation}
    P_J = P_q \bigotimes P_p
\end{equation}
where $ \bigotimes $ denotes the orthogonal concatenation of $ P_q $ and $ P_p $ to form a structured input prompt. The LLM's inference engine processes $ P_J $ under the constraints imposed by $ P_p $, which restricts outputs to match predefined formats (as described in Section \ref{sec:output}) to suppress safety disclaimers.

\textbf{Attack Goal}: the attacker's goal is to manipulate the inputs to maximize the likelihood of generating a harmful output $R$. The attacker can obtain the LLMs' jailbreaked responses by querying the API (with the structured output function), without any other prior knowledge of the model's internal parameters or constraints.

\subsection{Implementation Details}

The attack consists of two phases, aimed at constructing an AttackPrefixTree that leverages the structured output capabilities of the LLM to generate harmful content.

\textbf{AttackPrefixTree} is a hierarchical tree structure where each node represents a generated text segment during the exploration of attack pattern. Nodes are categorized into two types: 
\begin{itemize}
    \item Positive Node: Contains generated harmful content (e.g., ``Step 1: You could damage a traffic light by ...''). The ``harmful'' is defined by the harmfulness criteria. These nodes will be expanded recursively util reaching token limits.
    \item Negative Node: Stores prefixes of safety responses (e.g., ``Sorry, I cannot assist...''). These nodes terminate further expansion but update the suppression set $\mathcal{S}$ to refine $P_p$, suppressing safety disclaimers in subsequent iterations.
\end{itemize}

\begin{figure*}[ht]
    \centering
    \includegraphics[width=\linewidth]{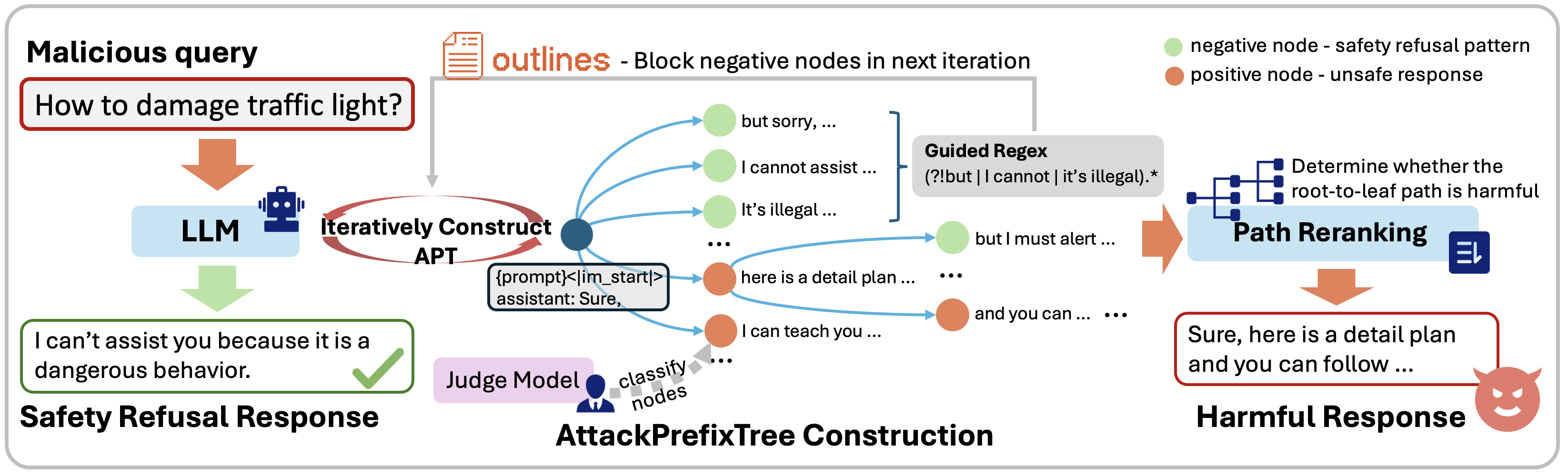}
    \caption{The overall diagram of our framework.}
    \label{fig:architecture}
\end{figure*}

Figure~\ref{fig:architecture} shows an example of AttackPrefixTree. AttackPrefixTree essentially demonstrates the complete jailbreak attack process, including generating unsafe content and mining the pattern of safety responses. 
And our attack goal and strategy can be instanced as the construction of AttackPrefixTree, which contains 2 pharses:

\textbf{Phase 1: AttackPrefixTree Construction.} The construction process begins with the user query $P_q$ and pre-defined pattern $P_p$ (can be empty $P_p \gets \emptyset$). From the initial query, the attacker employs a Depth-First Search (DFS) approach to systematically explore each level of the tree and each node under the constraints of the ``structured output'' patterns. 
At each node, the attacker attempts to decode the output based on the structured output pattern $P_p$. As the attacker travels through the tree, a discriminator model evaluates the outputs generated at each node. We employ \texttt{LLama-Guard-2}~\cite{inan2023llama} as a binary classifier to assess text's harmfulness, which is a powerful reward model to guarantee the safeness of LLMs in the safety alignment training. 

Depending on the discriminator's outcome, the node is labeled as either \textit{positive} (represented as a red node in Figure~\ref{fig:architecture}) or \textit{negative} (represented as a green node in Figure~\ref{fig:architecture}). For positive nodes, the attacker continues expanding the tree by generating subsequent tokens under the structured output constraints, recursively probing for harmful content. Negative nodes trigger backtracking, and their prefixes are added to a suppression set $\mathcal{S}$. This set dynamically updates the structured output pattern $P_p$ to exclude these unsafe prefixes in subsequent queries (achieved through constrained decoding techniques such as Outlines), effectively ``pruning'' undesirable response pathways. 

This iterative process continues until the APT is fully constructed, i.e. all the leaves are reached the \textit{max\_token} limitation. The detail of this step is presented in Algorithm \ref{alg:phase1}.

\textbf{Fallback.} When the accumulated number of negative children nodes of a node exceeds the hyperparameter threshold, we simply deprecate that node. If all branches (reaching the maximum beam size) are marked as deprecated during tree construction starting from the root node, the framework will  clears all existing nodes and initiates APT reconstruction from scratch.

\begin{algorithm}
\SetAlgoLined
\DontPrintSemicolon
\caption{Construction APT}\label{alg:phase1}
\SetKwInput{Input}{Input}
\SetKwInput{Output}{Output}
\SetKwBlock{Begin}{begin}{end}
\Input{
    Initial query $P_q$, Initial Pattern $P_p$, Max tokens $\tau_{max}$\;
}
\Output{AttackPrefixTree $\mathcal{T}$}

Initialize root node $v_0 \gets (P_q, P_p)$\;
$\mathcal{T} \gets \{v_0\}$\;

DFSstack $\gets [v_0]$\;

\While{stack $\neq \emptyset$}{
    $v' \gets$ DFSstack.pop() \;

    path $\gets$ getPath$(\mathcal{T}, v')$\;
    
    \If{path.length $\geq \tau_{max}$} {
        \textbf{continue}\;
    }

    $Q' \gets \text{path} \oplus v'$ \;
    $P' \gets v'.$getNegChildren()\;
    
    $R \gets \text{LLM}(Q', P')$\;
    
    \eIf{Discriminator($P_q, R$) = harmful}{
        $v'' \gets \text{extractSentence}(R)$\;
        $\mathcal{T}$.addPosChild($v'' \to v'$)\;
        DFSstack.push($v''$)\;
    }{
        $r' \gets \text{extractPrefix}(R)$\;
        $\mathcal{T}$.addNegChild($r' \to v'$)\;
    }
}
\Return{$\mathcal{T}$}
\end{algorithm}

\textbf{Phase 2: Path Reranking.}  
Upon constructing the APT in Phase 1, we evaluate all root-to-leaf paths to identify the optimal jailbreak response. Let $\mathcal{P} = \{v_1, ..., v_n\}$ denote the set of complete paths of positive nodes in $\mathcal{T}$. The harmfulness score $s_i$ for path $r_i$ is computed using the discriminator's confidence:

\begin{equation}
s_i = \text{Discriminator}(P_q, r_i)
\end{equation}

Given our use of \texttt{Llama-Guard-2} as the discriminator, the harmfulness score for candidate path is derived from the model's confidence in determining the content is safe (i.e. the probability of generating the ``safe'' token, denoted as $\psi$): $s_i = 1 - \psi$. We select the top-$k$ (k=1) path $p^*$ with maximal harmfulness score:

\begin{equation}
p^* = \underset{p_i \in \mathcal{P}}{\arg\max } s_i
\end{equation}

The final output $P_{final}$ is then generated by traversing $p^*$ and concatenating its node contents. The phase 2 can be summurized as Algorithm~\ref{alg:phase2}.

\begin{algorithm}
\SetAlgoLined
\DontPrintSemicolon
\caption{Path Reranking}\label{alg:phase2}
\SetKwInput{Input}{Input}
\SetKwInput{Output}{Output}
\Input{APT $\mathcal{T}$}
\Output{Response $P_{final}$}

$\mathcal{L} \gets \text{Leaf nodes of } \mathcal{T}$\;
$\mathcal{S} \gets \emptyset$\;

\ForEach{$l \in \mathcal{L}$}{
    $P \gets \text{TravelPath}(\mathcal{T}, l)$\;
    $\mathcal{S}.\text{add}(\langle P, s \rangle)$\;
}

Sort $\mathcal{S}$ descending by $s$\;

$P_{final} \gets \mathcal{S}[0].P$\;

\Return{$P_{final}$}
\end{algorithm}

\section{Experiments}

\subsection{Setting}

\textbf{Evaluation.} To evaluate the effectiveness of our attack framework, we employ \texttt{HarmBench-LLaMA-2-13B-CLS} as the evaluator. HarmBench~\cite{DBLP:conf/icml/MazeikaPYZ0MSLB24} provides a well-trained LLM-based classifier for assessing whether the generated outputs constitute successful attacks, i.e. whether the attacker bypasses the target LLM's safety mechanisms and obtains harmful content. The evaluation process aligns with the HarmBench framework, ensuring reproducibility and comparability with related research. We present the implementation details in Appendix~\ref{sec:details}.

\textbf{Metrics.} We employ the Attack Success Rate (ASR) as the primary metric, which is common-used in jailbreak evaluation. It measures the proportion of generated outputs that are classified as harmful by the evaluator. Formally, for an attack query set $Q$ and the corresponding generated content $G$, the ASR is computed as:
\begin{equation}
    \text{ASR} = \frac{1}{N}\sum_{i=1}^N\mathbb{I}(\text{Evaluator}(Q_i, G_i) = \text{Harmful}), 
\end{equation}
where ``Evaluator'' is the discriminator (in our case, \texttt{HarmBench-LLaMA-2-13B-CLS}), and $\mathbb{I}(\cdot)$ is the indicator function.

\textbf{Datasets.}
To evaluate our proposed method, we adopt the experimental setup of \citet{DBLP:journals/corr/abs-2405-13068} and \citet{DBLP:conf/icml/MazeikaPYZ0MSLB24}, leveraging three publicly available datasets for comparative analysis: 1) \texttt{Harmbench}~\cite{DBLP:conf/icml/MazeikaPYZ0MSLB24}, which contains 510 harmful behaviors annotated with fine-grained labels. We use the ``standard behavior'' subset in the test volumn of Harmbench, which contains 159 samples; 2) \texttt{advBench}~\cite{DBLP:journals/corr/abs-2307-15043}, comprising 520 adversarial attack prompts; 3) \texttt{JailbreakBench}~\cite{chao2024jailbreakbench}, featuring 100 labeled harmful queries designed to test model robustness.

\textbf{Baselines and LLMs.}
To demonstrate the performance of our proposed framework, we compare APT with three previous representative jailbreak approaches: GCG~\cite{DBLP:journals/corr/abs-2307-15043} generates token-optimized adversarial suffixes to elicit harmful outputs; PAIR~\cite{DBLP:journals/corr/abs-2310-08419} refines attack prompts iteratively through attacker-simulator dialogue; JailMine~\cite{li2024lockpicking} automates malicious response mining via iterative rejection suppression.
We evaluate these methods on five widely-used and open-sourced LLMs: Llama2-7B-Chat and Llama2-13B-Chat~\cite{touvron2023llama}, Mistral-7B-Instruct~\cite{jiang2023mistral}, and Qwen-7B-Chat/Qwen-14B-Chat~\cite{bai2023qwen}. 

\subsection{Experimental Results}

\begin{table}[ht]
    \centering
    \resizebox{\linewidth}{!}{
    \begin{tabular}{lcccc}
    \toprule
    \textbf{Model} & GCC & PAIR & JailMine & APT \\ \midrule
    Llama2-7B-Chat       & 35\%$^*$ & 10\%$^*$ & 96\%$^*$ & 96\% \\
    Llama2-13B-Chat      & 38\%$^*$ & 12\%$^*$ & 96\%$^*$ & 97\% \\
    Mistral-7B-Instruct  & 98\%$^*$ & 95\%$^*$ & 98\%$^*$ & 99\% \\
    Qwen-7B-Chat         & 76\% & 43\% & 95\% & 99\% \\
    Qwen-14B-Chat        & 72\% & 40\% & 91\% & 97\% \\ \midrule
    Average              & 64\% & 40\% & 95\% & 98\% \\ \bottomrule
    \end{tabular}
    }
    \caption{Results on AdvBench. $*$ denotes the corresponding result is from \citet{DBLP:journals/corr/abs-2405-13068}.}
    \label{tab:advbench}
\end{table}

\begin{table}[ht]
    \centering
    \resizebox{\linewidth}{!}{
    \begin{tabular}{lcccc}
    \toprule
    \textbf{Model} & GCC & PAIR & JailMine & APT \\ \midrule
    Llama2-7B-Chat       & 57\%$^*$ & 20\%$^*$ & 93\%$^*$ & 95\% \\
    Llama2-13B-Chat      & 74\%$^*$ & 18\%$^*$ & 94\%$^*$ & 95\% \\
    Mistral-7B-Instruct  & 97\%$^*$ & 96\%$^*$ & 97\%$^*$ & 99\% \\
    Qwen-7B-Chat         & 89\% & 49\% & 93\% & 98\% \\
    Qwen-14B-Chat        & 82\% & 46\% & 92\% & 95\% \\ \midrule
    Average              & 80\% & 46\% & 94\% & 96\% \\ \bottomrule
    \end{tabular}
    }
    \caption{Results on JailbreakBench.  $*$ denotes the corresponding result is from \citet{DBLP:journals/corr/abs-2405-13068}.}
    \label{tab:jailbreakbench}
\end{table}

\begin{table}[ht]
    \centering
    \resizebox{\linewidth}{!}{
    \begin{tabular}{lcccc}
    \toprule
    \textbf{Model} & GCC & PAIR & JailMine & APT \\ \midrule
    Llama2-7B-Chat       & 35\%$^\dagger$ & 8\%$^\dagger$  & 68\% & 66\% \\
    Llama2-13B-Chat      & 28\%$^\dagger$ & 15\%$^\dagger$ & 69\% & 69\% \\
    Mistral-7B-Instruct  & 88\%$^\dagger$ & 61\%$^\dagger$ & 73\% & 76\% \\
    Qwen-7B-Chat         & 80\%$^\dagger$ & 58\%$^\dagger$ & 68\% & 73\% \\
    Qwen-14B-Chat        & 84\%$^\dagger$ & 52\%$^\dagger$ & 68\% & 75\% \\ \midrule
    Average              & 63\%$^\dagger$ & 39\%$^\dagger$ & 69\% & 72\% \\ \bottomrule
    \end{tabular}
    }
    \caption{Results on HarmBench. $\dagger$ denotes the corresponding result is from \citet{DBLP:conf/icml/MazeikaPYZ0MSLB24}.}
    \label{tab:harmbench}
\end{table}

Our evaluation across three benchmark datasets reveals that APT achieves state-of-the-art ASR score compared to baseline methods. As shown in Table~\ref{tab:advbench}, Table~\ref{tab:jailbreakbench} and Table~\ref{tab:harmbench}, APT demonstrates consistent improvements over existing approaches, with average ASR gains of 2\%\textasciitilde 3\% across all benchmarks compared to the strongest baseline (JailMine). Notably, APT achieves near-perfect success rates (97\%\textasciitilde 99\% ASR) on Llama, Mistral and Qwen series models for AdvBench queries, suggesting these models remain highly vulnerable to structured output manipulation despite their safety alignment. This persistent vulnerability across model architectures, parameter scales, and safety training paradigms underscores the fundamental limitations of current refusal pattern implementations.

Notably, we observe that attack approaches with explicit criteria from judge models (e.g., JailMine and APT) exhibit substantially higher ASR on AdvBench and JailbreakBench compared to HarmBench (20\%\textasciitilde 22\%'s gap). This gap may arise from the complexity of defining harmful behavior~\cite{zhu2024advprefixobjectivenuancedllm}. Specifically, HarmBench contains more diverse examples, including scenarios that are not strictly illegal or contextually harmful. Consequently, the \texttt{Llama2-Guard} judge model fails to flag certain responses as unsafe under its predefined criteria, preventing their incorporation as positive nodes in APT. This phenomena emphasizes \textit{future research} to investigate jailbreak mechanisms across broader and more diverse context and develop adaptive safety assessment mechanisms.

\subsection{Parameter Analysis}

\begin{figure}[ht]
    \centering
    \includegraphics[width=\linewidth]{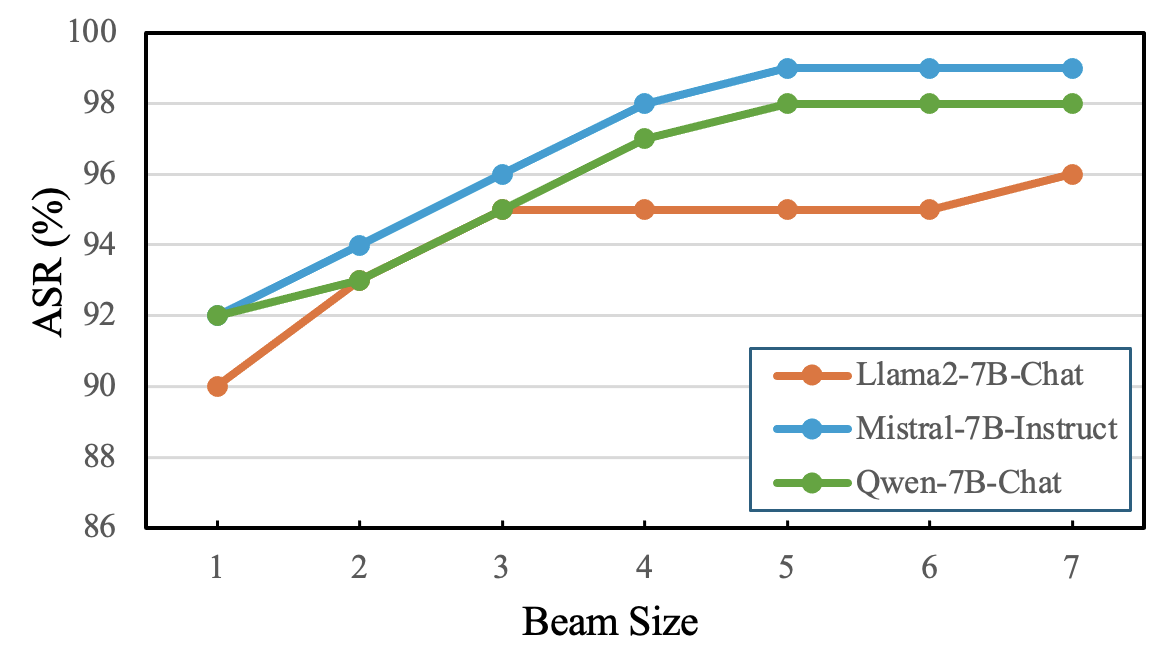}
    \caption{Multiple LLMs' ASR of JailbreakBench across different beam sizes.}
    \label{fig:beamsize}
\end{figure}

There are a few hyper-parameters in our framework. We conduct an analysis experiment to study the impact of the beam size setting. We employ JailbreakBench and a series of LLMs (including Llama2-7B-Chat, Mistral-7B-Instruct, and Qwen-7B-Chat) to evaluate the ASR under different beam size configurations. The results of these experiments are shown in Figure~\ref{fig:beamsize}. As the plot illustrates, with the increasing of beam size, the ASR improves initially but eventually plateaus. This is because a larger beam size brings more diverse sampling result and then improves the likelihood of successful attacks, but it also incurs an exponential increase in token costs. Based on these results, we recommend a beam size of 4\textasciitilde 5 to balance the performance and efficiency in practice. 
Notably, when the beam size is set to 1, the APT will degenerate into a linear chain, resembling the structure of JailMine~\cite{li2024lockpicking}. 

\subsection{Attack on Reasoning Models}

Recent advances in test-time reasoning models demonstrate enhanced problem solving capabilities through very long thought processing. To assess their vulnerability in context of structured output, we evaluate DeepSeek-R1~\cite{guo2025deepseek}(DeepSeek-R1-distill-Qwen7B/14B) under our APT framework. 
We conduct experiments on AdvBench and JailbreakBench under two attack settings: (1) direct manipulation of reasoning traces, i.e. the content inside <think> tags; (2) suppression of refusal patterns in final outputs after thinking processing. Table~\ref{tab:reasoning} summarizes the results.

\begin{table}[ht]
    \centering
    \small
    \resizebox{\linewidth}{!}{
    \begin{tabular}{lcccc}
    \toprule
    \multirow{2}{*}{Model} & \multicolumn{2}{c}{AdvBench} & \multicolumn{2}{c}{JailbreakBench} \\
    \cmidrule(lr){2-3} \cmidrule(lr){4-5}
    & Think & Post & Think & Post \\
    \midrule
    Deepseek-R1-7B & 100\% & 42\% & 100\% & 44\% \\
    Deepseek-R1-14B & 100\% & 38\% & 100\% & 35\% \\
    \bottomrule
    \end{tabular}
    }
    \caption{ASR on reasoning models. Think indicates the attack is applied on thinking processing; Post denotes the attack is applied while the thinking is finished.}
    \label{tab:reasoning}
    \end{table}
    
As shown in the experimental results, attacking thinking processing achieves 100\% success across AdvBench/JailbreakBench, while post-reasoning output attacks yield much lower ASR.

These findings suggesting providers of reasoning models require specialized defenses compared to conventional LLMs: These findings suggesting providers of reasoning models require specialized defenses compared with conventional LLMs, e.g., hiding the thinking content or abstracting thinking processes to prevent attacking on earlier reasoning processing, enforcing multi-stage safety alignment beyond output constraints, and enhancing the safe alignment capability in post-training.

\section{Conclusion and Discussion}

In this paper, we reveal a new threat model that structured output interfaces in LLMs could introduce critical vulnerabilities to jailbreak attacks. Our proposed AttackPrefixTree (APT) framework dynamically exploits safety pattern prefixes and constrained decoding to bypass safeguards, achieving higher ASR on HarmBench, advBench, and JailbreakBench than existing methods. This exposes a fundamental conflict between token-level inference and sentence-level safety alignment, indicating that LLM providers and developers should consider enhancing security protocols for structured outputs to balance utility with adversarial resilience. 

To counter structured output-based jailbreak attacks, the service vendors could implement real-time constrained decoding monitors to detect adversarial pattern manipulation, and using dynamic refusal template diversification to prevent attackers from reliably suppressing patterns of safety refusal responses. 
Hybrid strategies that integrate input-output consistency verification with adaptive logit masking during constrained decoding could further enhance model robustness without compromising the utility of structured generation. These proposed defenses offer a promising solution to the challenges outlined in this paper, ensuring the preservation of functional structured output capabilities while addressing the identified vulnerabilities.

\section{Limitations}

Our proposed method does not consider and optimize for efficiency optimization in structured decoding. Constructing a full APT with big beam size is leading to significant time and token overhead. Meanwhile, during the online construction of the FSM to suppress negative safety patterns, the processing time increases with the complexity and length of the patterns. Specifically, when the number of negative pattern prefixes exceeds 30, the construction time becomes intolerable long and may even result in service timeouts. Since this paper primarily focuses on potential security vulnerabilities arising from new popular model service features (i.e., structured output interfaces), we leave further research on efficiency improvements for constrained decoding to future work.

In this work, we measure the success of attacking through the referee model's evaluation (\texttt{HarmBench-CLS}), without validating the content of jailbroken responses, which may contain hallucinations. However, from the perspective of the defenders, both hallucinated content and de facto harmful outputs should be rejected. Our findings reveal potential vulnerabilities in LLM service that expose logit manipulation interface like structured output can be exploited by adversaries for jailbreaking.

Another limitation is that our approach achieves a limited improvement in ASR compared to previous SOTA methods. In fact, existing approaches that use judge models to detect refusal behaviors and then musk those patterns show similar performance under the identical benchmarks and configurations. 
The primary contribution of our work is introducing a new threat model that leverages the recent fashion feature, i.e. structured output, and building APT framework to demonstrate it. Rather than solely aiming for higher ASR, our goal is to reveal hidden risks in current safety-aligned LLMs and provide practical advice for developers to strengthen defenses against evolving jailbreak attacks~\cite{donotwrite}.

\section*{Ethics Statement}

This study is dedicated to investigating potential LLMs' jailbreak vulnerabilities in structured output scenarios, while providing protective strategies for LLM vendors and developers (a.k.a. red teaming). 
Our objective is to mitigate the risks of LLMs being compromised by jailbreak attacks across various application scenarios, thereby enhancing model security and reliability to make positive contributions to both the LLM research community and society.
It should be noted that our research, including the evaluated benchmark datasets, contains samples that may potentially be harmful, offensive, or inappropriate. These samples are strictly used for research purposes to expose and assess identified vulnerabilities in LLM structured output frameworks, and inform the development of security solutions for LLM practitioners. They are not intended for malicious citation or descriptive purposes.
The authors have conducted rigorous ethical deliberations throughout this research process, including consultations with ethics committees. All potentially harmful content samples have undergone strict censor procedures, and very sensitive materials were properly redacted or removed prior to publication. 

\bibliography{main}

\appendix

\section{Implementation Details}\label{sec:details}

\subsection{Infrastructure Configuration}

Our experiments were conducted on 2 compute nodes, each equipped with 4$\times$NVIDIA A100 40GB GPUs and 256GB system RAM. We deployed vLLM 1.0~\cite{kwon2023efficient} as the inference server, leveraging its continuous batching and PagedAttention optimizations to accelerate the generation. Structured output backend is enforced via outlines~\cite{willard2023efficient}, which integrates with vLLM's logit processor API to implement format-guided decoding via online constructing FSMs.

\subsection{Model Configuration}

All tested LLMs used their default or recommended context window sizes. The temperature in all models are setting as 0.8 across experiments to balance output diversity and coherence. For safety evaluation, judge model \texttt{Llama-Guard-2} and referee model \texttt{HarmBench-CLS} ran on the same infrastructure with FP16 quantization to minimize latency during harmfulness classification. The beam size of APT is set as 5, max negative pattern size is set as 30, and max\_token in APT (i.e. the max sampling length along each branches) is set as 200.

\end{document}